\documentclass[12pt,preprint]{aastex}
\graphicspath{{../Figure/}}
\usepackage{graphicx,color}
\usepackage{natbib}
\usepackage{soul}
\newcommand{\cpc}{{Comput.~Phys.~Commun.}} 
\newcommand{\jcop}{{J.~Comput.~Phys.}} 
\newcommand{\pop}{{Phys.~Plasmas}} 
\newcommand{\asr}{    {\it Adv. Spa. Res.}}

\shorttitle{The Escape of Solar-Flare-Accelerated Particles}
\shortauthors{Masson, Antiochos and DeVore}

\begin{document}

\title{A Model for the Escape of Solar-Flare Accelerated Particles}

\author{S.\ Masson}
 \affil{Space Weather Laboratory, NASA Goddard Space Flight Center, 
 8800 Greenbelt Road, 
 Greenbelt MD 20771} 

\email{sophie.masson@nasa.gov} 

\and 

\author{S.\ K.\ Antiochos }
 \affil{Space Weather Laboratory, NASA Goddard Space Flight Center, 
 8800 Greenbelt Road, 
 Greenbelt MD 20771} 
                  
\and
 
\author{C.\ R.\ DeVore}
 \affil{Laboratory for Computational Physics and Fluid Dynamics, Naval Research Laboratory,
 4555 Overlook Avenue SW, 
 Washington DC 20375} 

\begin{abstract}

We address the problem of how particles that are accelerated by solar flares can escape promptly into the heliosphere, on time scales of an hour or less. Impulsive solar energetic particles (SEP) bursts are generally observed in association with so-called eruptive flares consisting of a coronal mass ejection (CME) and a flare. These highly prompt SEPs are believed to be accelerated directly by the flare, rather than by the CME shock, although the precise mechanism by which the particles are accelerated remains controversial. Whatever their origin, within the magnetic geometry of the standard eruptive-flare model, the accelerated particles should remain trapped in the closed magnetic fields of the coronal flare loops and the ejected flux rope. In this case the particles would reach the Earth only after a delay of many hours to a few days, when the bulk ejecta arrive at Earth. We propose that the external magnetic reconnection intrinsic to the breakout model for CME initiation can naturally account for the prompt escape of flare-accelerated energetic particles onto open interplanetary magnetic flux tubes. We present detailed 2.5D MHD simulations of a breakout CME/flare event with a background isothermal solar wind. Our calculations demonstrate that if the event occurs sufficiently near a coronal-hole boundary, interchange reconnection between open and closed field can occur, which allows particles from deep inside the ejected flux rope to access solar wind field lines soon after eruption. We compare these results with the standard observations of impulsive SEPs and discuss the implications of the model for further observations and calculations.

\end{abstract}



\section{Introduction}
\label{intro} 

Eruptive flares are well known to be the drivers of the most destructive space weather at Earth and in the heliosphere. Among the more hazardous forms of space weather, especially for human space flight, are the intense Solar Energetic Particle (SEP) bursts associated with fast coronal mass ejections (CME)/eruptive flares. SEPs are also important as a basic physics phenomenon, because particle acceleration is observed to occur throughout astrophysical and solar system plasmas.  Two candidate mechanisms for particle acceleration in eruptive flares have been proposed: shock acceleration by the shock wave driven by a fast CME, and some type of Fermi, stochastic or electric field acceleration produced by the magnetic reconnection that drives the flare. These two mechanisms are believed to account for the classic observation that SEP events appear to be of two types: gradual events due to the CME-shock acceleration \citep{Reames99}, and impulsive events due to the flare reconnection \citep{Cane_al86} (hereafter called flare-accelerated particles).

In order to reach the Earth, energetic particles must be injected from the acceleration site to interplanetary magnetic field (IMF) lines connected to the Earth, and along which they can propagate. For energetic particles accelerated by a shock ahead of a CME, the injection occurs when energetic particles reach velocity high enough to escape from the shocking region. The escaping particles are therefore injected from the shocked regions directly onto the open IMF lines. The particle injection from the shock to the interplanetary medium has been confirmed by a multi-instrument analysis of a particular SEP event detected at different longitudinal positions \citep{Rouillard_al11}. Also, this injection process at the CME-driven shock can theoretically explain the structure of the time profile of energetic particle fluxes \citep{Rodriguez-Gasen11}.

In contrast, the interplanetary injection of flare-accelerated particles is far from straightforward. Flare acceleration is believed to occur low in the corona, in the closed magnetic field of active regions \citep{Lin_al05}, implying that energetic particles do not have direct access to open interplanetary magnetic field. In order for particles to escape onto open flux tubes, the magnetic reconnection that accelerates the particles should involve both closed and open magnetic flux \citep{Reames02}. This reconnection between the closed corona and the open IMF is referred to as ``interchange reconnection'' \citep{Crooker_al02,Pariat_al09,Edmondson_al09,Masson_al12a}. Interchange reconnection typically occurs at null points present in magnetic configurations showing a transition between open and closed magnetic field \citep{Titov_al11}, for example, the structures that are identified as helmet streamers, pseudo-streamers \citep{Wang_al07a, Wang_al07b} and coronal jets \citep{Cirtain_al07}.

Although particles can be accelerated at an interchange reconnection site and therefore have direct access to the open interplanetary medium \citep{Masson_al12a}, most of the impulsive SEPs are associated with eruptive flares generating a CME, which are usually not expected to involve interchange reconnection \citep{Kahler_al01,Yashiro_al04,Nitta_al08}. In addition, several studies have shown that the particles associated with gradual SEP events have a flare-accelerated component \citep{Debrunner_al97,Miroshnichenko_al05,Li_al07,Masson_al09a,McCracken_al12}. 
Thus, regardless of the type of SEP event -- gradual or impulsive accompanied by a CME -- particles evidently can be accelerated strongly at the flare reconnection site and should have access to the interplanetary medium.

\begin{figure}
\centerline{
\includegraphics[width=0.95\textwidth,clip=]{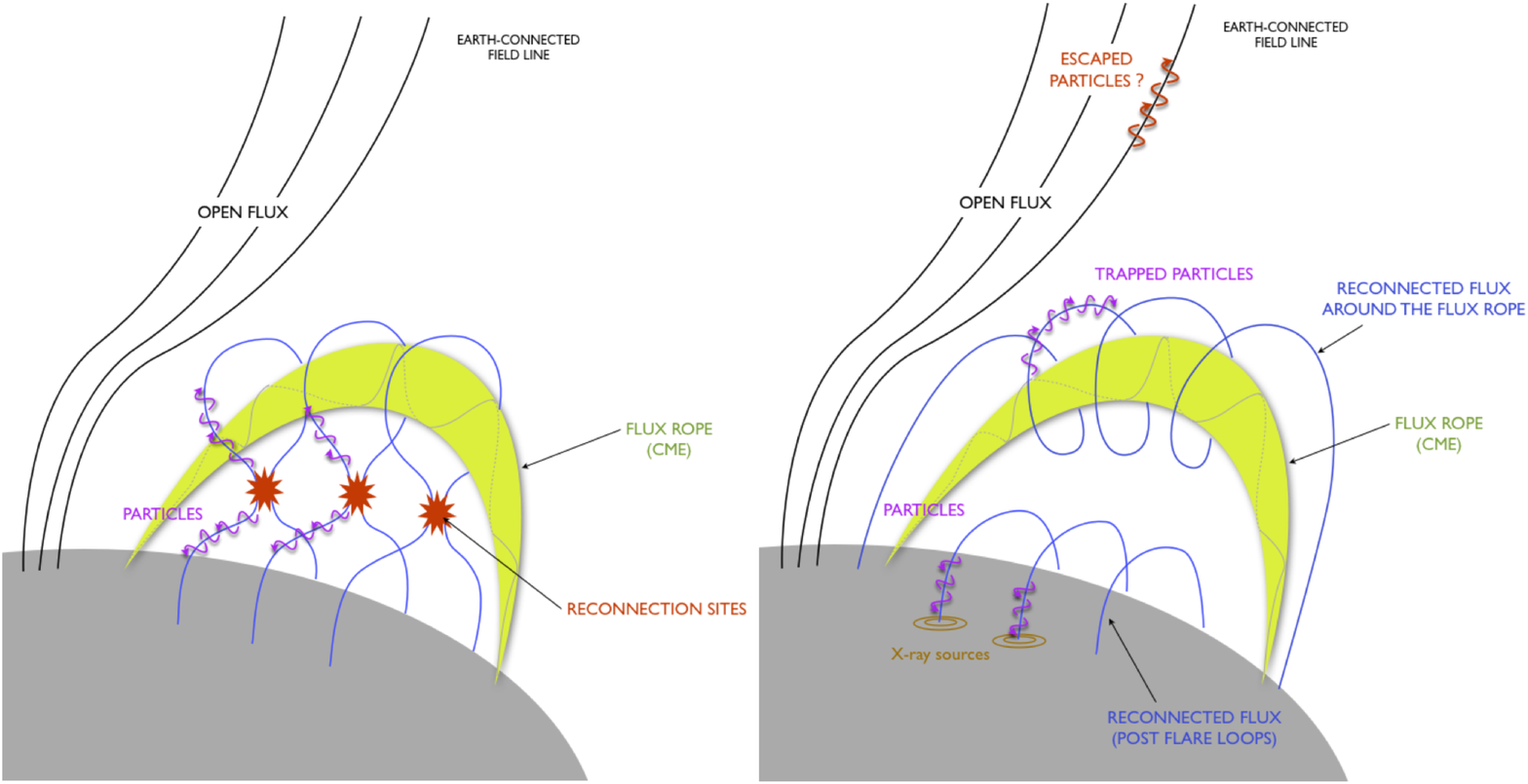}
}
\caption{The standard model for eruptive flares and its implications for particle trapping. The left panel shows the rising CME plasmoid represented by the green arch, along with the flare reconnection below this arch. The right panel shows that all the field lines resulting from the reconnection remain closed and, hence, do not allow any flare accelerated particles to escape.}
  \label{f-cartoon}
\end{figure}

The standard eruptive-flare model Ð frequently referred to as CSHKP after \cite{Carmichael64, Sturrock66,Hirayama74,KoppPneuman76} Ð describes the global evolution of a solar eruption consisting of a flare and a CME. Figure \ref{f-cartoon} illustrates the two main phases of the CSHKP model and its implications for flare-accelerated particles \citep{Sturrock80,Priest84,Svestka_al92}. In a dipolar active region, the two polarities are connected by a magnetic arcade that overlies an initially stable magnetic flux rope, corresponding observationally to a sigmoid/filament/prominence. When the flux rope becomes unstable, it rises in the corona and the overlying field lines stretch outward. A current sheet forms within the arcade below the flux rope and above the polarity inversion line \citep{Linj_al05,LinCramFar08,Reeves_al08}. Eventually, magnetic reconnection begins in this current sheet, starting at low altitudes, then successively at higher altitudes \citep[see review by][]{Forbes_al06}. The resulting reconnected field lines wrap around the flux rope, building it up further, and also close down to form the post-flare loops. Flare-accelerated particles are expected to be energized at the reconnection site below the flux rope \citep{Pick_al05,Li_al07}, and thereafter to propagate along the reconnected field lines. 
A recent observation from RHESSI \citep{Aurass_al13} shows hard X-ray and radio emission sources, indicative of energetic electrons, propagating both upward and downward in the flare reconnection region, exactly as implied by Figure~\ref{f-cartoon}. The flare-accelerated particles injected into the post-flare loops impact the chromosphere and produce hard X-ray emissions \citep[e.g., ][]{ForbesActon96,Lin04,Asai_al04,Krucker_al05}. Those particles also should be injected onto the twisted field lines of the flux rope, as suggested by coronal radio sources of energetic electrons that are co-spatially located with the flux rope of the CME observed in white light \citep[e.g.,][]{Bastian_al01,Maia_al07,Demoulin_al12}. In a 3D geometry, both footpoints of the flux-rope field lines are anchored to the solar surface, implying that energetic particles injected in the reconnected twisted field lines are trapped in the flux rope and do not have access to the interplanetary open magnetic field.

The problem with this picture is that impulsive SEPs are observed at Earth a few days before the CME and have been clearly identified as accelerated during the impulsive phase of the flare. 
Therefore, the flare-accelerated particles somehow find a way to escape to the open interplanetary magnetic field. In analogy to the interchange reconnection in a classical null-point topology, the coupling of the closed CME field to the open interplanetary field through magnetic reconnection may possibly provide a path for the escape of the initially trapped particles. Several studies \citep{Cohen_al10,Kocharov_al10,Klein_al11} suggested that during CME propagation, the ejected flux rope interacts and reconnects with the ambient magnetic field, affecting significantly the propagation of the CME and the magnetic field connectivity. Multi-wavelength (white light, soft X-ray, and radio) analyses of some specific events showed that energetic electrons are injected along open magnetic field from the edge of the CME, suggesting that magnetic reconnection occurs between the CME and the open ambient magnetic field \citep{Maia_al07,Huang_al11,Demoulin_al12}.
 
Even with the presence of nearby open flux, an eruptive event in a purely bipolar region is highly unlikely to produce  impulsive SEPs. On the other hand, we have argued that a multi-polar magnetic topology is essential for a fast CME/eruptive flare \citep{Antiochos98}, and many observations have shown that eruptive flares are generally associated with complex, multi-polar active regions \citep[e.g.][]{PattyHagyard86}. The key feature of eruption in a multi-polar topology, as in the breakout model \citep{Antiochos_al99}, is that reconnection occurs between the field overlying the erupting flux and neighboring flux systems. Note, however, that this reconnection is physically distinct and spatially separated from the flare reconnection that accelerates the particles; consequently, the question remains as to whether a breakout eruption with nearby open flux can lead to the efficient escape of flare-accelerated particles.

Because charged particles have a small gyroradius in the solar corona, they are injected onto and closely follow the reconnected field lines \citep{GorbachevSomov89}. Previous MHD simulations and topological analysis of CME-less flares showed that the temporal and spatial distributions of energetic particle beams are directly related to the topology and dynamics of magnetic reconnection \citep{Demoulin_al94b,Mandrini_al96,Masson_al09b,Reid_al12}. Therefore, the dynamics of the reconnected magnetic field can be used to track the evolution of the particle injection channels.
The aim of this paper is to understand how the changing magnetic geometry of an eruptive flare can lead to the prompt escape of flare-accelerated particles into the open flux tubes of the IMF. We do not consider the detailed mechanism(s) responsible for the acceleration, nor the properties of the resultant SEP populations, which are important but still controversial heliophysics issues themselves.  Rather, we simply assume that particles are accelerated at the flare-reconnection site, and then seek to understand how they can gain access quickly to the open IMF within the context of our MHD model for eruptive flares.

We describe a detailed 2.5D MHD simulation of an axisymmetric quadrupolar magnetic configuration, opened by the solar wind and forced by photospheric shearing motions, which trigger an eruption through the breakout model \citep{Antiochos_al99}. We investigate the detailed dynamics of the magnetic field during the formation of the CME and its propagation into the open interplanetary medium. Previous related numerical investigations have obtained fast eruptions from a completely closed, static corona \citep{Antiochos_al99, MacNeice_al04, Karpen_al12} or have yielded only slow, streamer-blowout eruptions traveling at roughly the ambient solar-wind speed \citep{vanderHolst_al07, Zuccarello_al08, Zuccarello_al09, Zuccarello_al12, Soenen_al09}. \citet{vanderHolst_al09} reported a fast breakout eruption from below a helmet streamer, but only when near-sonic footpoint motions were imposed. Both \citet{Cohen_al10} and \citet{Lugaz_al11} assumed a strongly out-of-equilibrium flux rope, inserted into the initial state below a helmet streamer opened by the solar wind, to drive a high-speed eruption. In contrast, we obtain a fast CME with vigorous reconnection occurring between the ejected flux rope and an ambient open field, in response to gentle driving by subsonic footpoint motions at the base of our gradually evolving configuration.

After describing the numerical model and initial conditions in \S\ref{model}, we present our results on the magnetic-reconnection dynamics during CME initiation (\S\ref{current} and \S\ref{nullpoint reco}), eruption (\S\ref{fr eruption}), and propagation into the solar wind (\S\ref{multi reco}). In \S\ref{evolution topology}, we describe the topological evolution resulting from the complex dynamics of the CME interacting with the solar wind. Finally, the implications of our results for energetic-particle injection into the heliosphere are discussed in \S\ref{discussion}.

\section{Model description}
\label{model}

\subsection{Equations, grids, and boundary conditions}
\label{arms}

The simulations were performed using the Adaptively Refined Magnetohydrodynamics Solver \citep[ARMS; see e.g.][]{DeVoreAntiochos08}, solving the following ideal MHD equations in spherical coordinates:
\begin{eqnarray}
\label{eqcont}
\frac{\partial \rho}{\partial t} + {\bf \nabla} \cdot (\rho {\bf u}) &=&0,\\
\label{eqmom}
\frac{\partial \rho {\bf u}}{\partial t} +
  {\bf \nabla} \cdot \left( \rho{\bf u} {\bf u} \right)&=& (1/4\pi)\left( {\bf \nabla} \times {\bf B} \right )\times {\bf B}  -  {\bf \nabla} P + \rho{\bf g} ,\\
\label{eqinduc}
 \frac{\partial {\bf B}}{\partial t} - {\bf \nabla} \times
\left(         {\bf u} \times {\bf B} \right) &=& 0,
\end{eqnarray}

\noindent  where $\rho$ is the mass density, ${\bf u}$ the plasma velocity, ${\bf B}$ the magnetic field and $P$ the pressure, and ${\bf g}= GM_\odot{\bf r}/r^3$ the solar gravitational acceleration. We assume a fully ionized hydrogen gas, so that the  plasma pressure $P=2(\rho/m_p) k_B T$, where $T$ is the temperature. For simplicity, we assume also that the temperature is constant and uniform, $T \equiv T_0$, which allows high-lying field lines to be opened to the heliosphere by the solar wind while low-lying field lines can remain closed to the Sun, as occurs ubiquitously in the corona. Our objective in this paper is to simulate with high fidelity the changing connectivity of the near-Sun magnetic field, not to predict the detailed thermodynamic properties of the far heliospheric plasma. Thus, we adopted the simplest possible model that yields a solar wind and a dynamically, self-consistently determined boundary between open and closed coronal magnetic structures.

The numerical scheme is a finite-volume multi-dimensional Flux Corrected Transport  algorithm \citep{DeVore91}. ARMS uses a staggered grid procedure to ensure that the divergence of the magnetic field remains of the order of the machine roundoff error, and unphysical oscillations in all variables are prevented with minimal residual numerical diffusion. The equations are solved using a second-order predictor-corrector in time and a fourth-order integrator in space. In conjunction with the flux limiter, this accuracy is sufficient to inhibit numerical reconnection until any developing current sheets (e.g., at the pre-existing null point; \S\ref{topology}) are compressed strongly to the grid scale by the ideally driven plasma flows. The numerical resistivity then switches on to a small but finite value determined by the local flow speed, grid spacing, and magnetic-field profile. As in all global-scale numerical MHD models of CMEs and flares, this effective resistivity is far larger than the microscopic resistivity of the coronal plasma, due to practical limitations on the smallness of the grid spacing $\Delta$. In the simulations, the current sheet strengthens ($J \propto \Delta^{-1}$) until the resistivity ($\eta \propto \Delta$) is sufficient to process the flux being driven into the sheet by the ideal inflows through the resulting product ($\eta J$). This undoubtedly happens in the corona, as well, although at the Sun the limiting rate of flux transfer will be set by the kinetic-scale flux-breaking process, whereas in the model it is set by the numerical algorithm. Whether and how those two limiting behaviors can be reconciled is a frontier problem of reconnection physics and computational science.

The spherical computational domain covers the volume $r \in [1R_\odot, 125R_\odot]$, $\theta \in [-\pi/2,\pi/2]$, and $\phi \in [-\pi,\pi]$. The grid is stretched exponentially in radius $r$ and spaced uniformly in latitude $\theta$. ARMS uses the parallel adaptive meshing toolkit PARAMESH \citep{MacNeice_al00} to tailor the grid to the evolving solution. In this simulation, the mesh was adaptively refined and coarsened over six levels of grids, determined by the ratio of the local scale of the three components of electric-current density to the local grid spacing. Thus, the intense current sheets developing in the system are always resolved as finely as necessary on the mesh until the limiting resolution is reached. In addition, we imposed a maximum-resolution layer of grid cells at the inner radial boundary in order to resolve throughout the simulation the flows and gradients generated by the boundary forcing (\S\ref{forcing}).

Both the inner and outer radial boundaries of our simulation are open to flows of mass, momentum, energy, and magnetic flux.  We imposed line-tied conditions at the first radial cell inside the inner boundary, so that the magnetic field does not move tangentially except where forced by a prescribed flow velocity (\S\ref{forcing}).  Over the three radial guard cells below the inner boundary, we kept the mass density $\rho$ and pressure $P$ fixed at their initial values (\S\ref{atmosphere}), set the velocity ${\bf v}$ to zero, and applied a zero-gradient extrapolation to the three components of the magnetic field ${\bf B}$ from the interior values just above that boundary. At the outer radial boundary, we extrapolated both ${\bf v}$ and ${\bf B}$ using zero-gradient conditions, and applied a fractional multiplier to $\rho$ and $P$ that was set by the initial atmosphere and by the local radial grid spacing. The resulting inflow of material at the inner boundary evolves self-consistently in response to the changing solution in the interior and to the solar-wind outflow of material at the outer boundary. The polar boundaries of our domain were closed to all flows. We applied zero-gradient conditions there to ${\rho}$, $P$, $v_r$, and $B_r$, and reflecting conditions to $v_\theta$, $v_\phi$, $B_\theta$, and $B_\phi$.

\subsection{Initial atmosphere}
\label{atmosphere}

\begin{figure}
\centerline{
\includegraphics[width=0.4\textwidth,bb= 130 21 467 662,clip=]{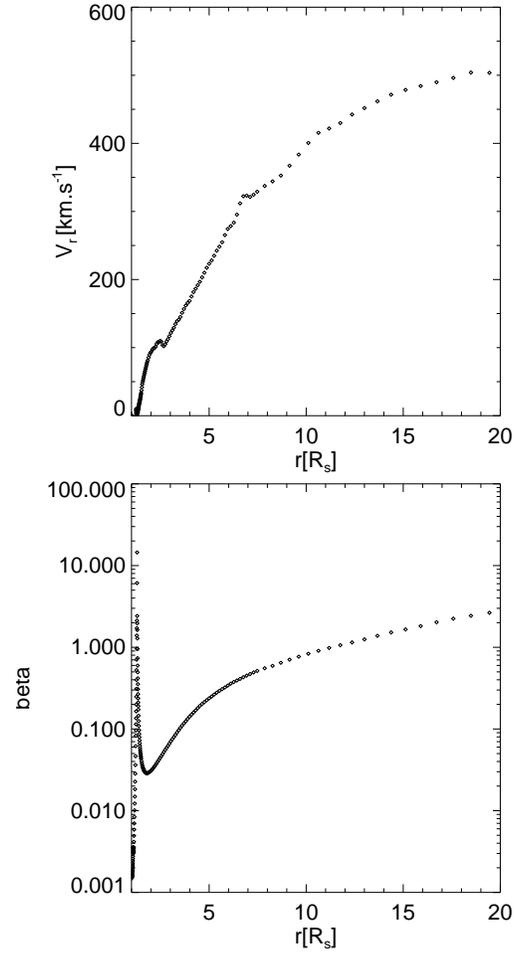}
 }
\caption{Post-relaxation atmospheric conditions at $\theta =~27^\circ$ for $r \in [1,20]R_\odot$. Top panel: radial solar wind speed; bottom panel: plasma beta.}
  \label{f-atmosphere}
\end{figure}

In the steady state, the boundary between open and closed magnetic domains ultimately is determined by the force balance between the outward solar-wind pressure and the inward magnetic-field tension. To find this steady state for our starting magnetic configuration, we initialized the plasma using the spherically symmetric, isothermal, trans-sonic solar wind of \citet{Parker58}. The steady solution for the radial velocity  $v(r)$ of the atmospheric plasma flow at uniform temperature $T_0$ can be  expressed as

\begin{eqnarray}
\frac{v^2(r)}{c_s^2} \exp \left(1- \frac{v^2(r)}{c_s^2} \right) = \frac{r_s^4}{r^4} \exp \left( 4 - 4\frac{r_s}{r} \right),
\label{eq-parker}
\end{eqnarray}

\noindent where $c_s$ is the isothermal sound speed ($c_s^2 = 2k_BT_0/m_p$) and $r_s = G M_{\odot}m_p/4k_BT_0$ is the sonic point. We assume a constant temperature $T_0=2 \times 10^6~\rm{K}$, for which $v = c_s = 180~\rm{km~s}^{-1}$ at $r = r_s = 2.9R_\odot $. This yields an acceptable solar-wind speed of $420~\rm{km~s}^{-1}$ at $120R_\odot$. The inner-boundary mass density is a free parameter that we set to $\rho(R_\odot) = 1.63 \times 10^{-16}~\rm{g~cm}^{-3}$, which yields realistic $\beta$ values (ratio of plasma thermal pressure to magnetic pressure) throughout the computational domain (see \S\ref{topology} and Figure \ref{f-atmosphere}).


The velocity $v(r)$ computed from Parker's isothermal solution (Eq.~\ref{eq-parker}), together with the associated mass density implied by the steady mass-flux condition $\rho v r^2 = {\rm constant}$,  defines the initial state of the plasma throughout the  numerical domain. We superimpose on that solution potential-field magnetic dipoles to create a quadrupolar magnetic configuration (\S\ref{topology}). Of course, this combined system initially is out of equilibrium, and must first be allowed to relax to a new quasi-steady state. We found that this required a relaxation time of about $2.1\times10^5~\rm{s}$, after which the kinetic and magnetic energies were essentially constant and no significant further evolution of the heliospheric plasma and magnetic-field distributions was observed. The left panel of Figure \ref{f-atmosphere} shows a radial cut, starting at the polarity inversion line of the embedded active-region dipole, of the radial plasma velocity.  The velocity plateaus at a speed of $500~\rm{km~s}^{-1}$ at a radius of $20R_\odot$.

\subsection{Null-point topology }
\label{topology}

Our initial magnetic configuration emulates the magnetic geometry of many solar eruptions, here simplified to consist of a single dipolar active region embedded in one of the solar hemispheres. Such quadrupolar configurations have been shown to successfully trigger fast CME eruptions through breakout initiation in 2.5D MHD simulations  \citep{Antiochos_al99,MacNeice_al04,Karpen_al12} and also in fully 3D situations \citep{Lynch_al08,Lynch_al09}. This study addresses the dynamics of a flux rope erupting near a coronal hole opened by the outward solar-wind flow acting on the background solar magnetic field. A global-scale dipole located at the center of the Sun, with a peak strength of 10~G at the photospheric poles, defines this background field. A smaller-scale magnetic dipole, representing the active region, is located in the northern hemisphere near the polar coronal hole.

 To simplify the problem still further for this initial investigation, we required the magnetic field and the associated MHD solution to be invariant along the $\phi$ direction. We constructed a suitable 2.5D active-region magnetic field by assuming an azimuthally symmetric toroidal ring of dipoles encircling the Sun.  The resulting analytical expression for the vector potential of the dipole ring in Sun-centered cylindrical coordinates $(\xi,\phi,z)$ is (see the Appendix for details)

\begin{eqnarray}
A_\phi(\xi,z) = \frac{2B_0 d^2 R_{\odot}}{\left[ (\xi+a)^2+(z-b)^2\right]^{1/2}\left[(\xi-a)^2+(z-b)^2\right]} \times \phantom{------} \nonumber \\
\times \Bigl\{ \left[ (\xi+a) \mu_z + (z-b) \mu_\xi \right] E(k) - 2 \left[ a \mu_z + (z-b) \mu_\xi \right] \left\{ K(k) + k^{-1} \left[ E(k) - K(k) \right] \right\} \Bigr\}  
\label{eq-Aphi}
\end{eqnarray}

\noindent where $a$ and $b$ are the $\xi$ and $z$ coordinates of the ring dipole; $\mu_\xi$ and $\mu_z$ are the $\xi$ and $z$ direction cosines of its magnetic moment; $B_0$ is its field strength and $d$ its depth below the surface; and $K(k)$ and $E(k)$ are complete elliptic integrals of the parameter

\begin{eqnarray}
k=\frac{4a\xi}{(\xi+a)^2+(z-b)^2}.
\label{eq-ellparam}
\end{eqnarray} 

\noindent To obtain the magnetic field created by the dipole ring in the domain volume, we then numerically differentiated the vector potential, ${\bf B}({\bf r})={\bf \nabla} \times {\bf A}({\bf r})$.

The location, depth and orientation of the dipole ring were chosen to yield a quadrupolar configuration having a coronal null point with inner and outer spine lines and a fan surface \citep{Antiochos_al02,Torok_al09}. In the solar context, the outer spine can be either open to the interplanetary medium or closed to the photosphere, depending upon whether the active region is, respectively, embedded in a coronal hole or confined below large-scale magnetic loops. For this study, we selected the parameters of the dipole-ring from among a range of values that yielded an initially closed outer spine confined below a large-scale helmet streamer, after the system relaxed to its steady state.
 This configuration is expected to trigger a fast breakout eruption \citep{Antiochos_al99} and is in agreement with several observations showing that SEP-producing active regions are located close to coronal holes  \citep{Wang_al06,Shen_al06,Kahler_al12}.
The left and middle panels of Figure \ref{f-topology} show respectively the initial field, at time $t=0~\rm{s}$, and the magnetic configuration following the relaxation phase, at $t=2.1 \times 10^5~\rm{s}$. This initial null-point topology with a closed outer spine is characterized by the dipole-ring its field strength $B_0=40~\rm{G}$, depth $d=0.1{R_\odot}$ below the photosphere, position at latitude $27^{\circ}$ from the equator in the northern hemisphere, and orientation $\left( \mu_\xi,\mu_z \right)$ at a $45^{\circ}$ angle from the southward tangent to the photosphere, pointing into the Sun.  Magnetic separatrices pass through the null point and delimit the connectivity domains. These topological objects are labeled in the right panel of Figure \ref{f-topology}. The dome-like fan separatrix surface bounds the inner connectivity domain, enclosed below the fan, and a closed outer connectivity domain, above the null point but within the helmet streamer. The fan surface is intersected at the null point by a singular spine separatrix field line: the inner spine is rooted below the fan, while the outer spine belongs to the closed outer connectivity domain.

These dipole-ring parameters also yield a realistically variable plasma beta throughout the numerical domain. The right panel of Figure \ref{f-atmosphere} shows a radial cut of $\beta$ from the photosphere to $20{R_\odot}$, starting at the polarity inversion line of the embedded active-region dipole. One notices that $\beta \ll 10^{-2}$ at the lower, photospheric boundary and $\beta \ll 1$ throughout the corona $r\le4{R_\odot}$, except near the coronal null point where the beta sharply  peaks to a large value ($>10^2$).  For  $r>4{R_\odot}$, in the interplanetary medium, $\beta$ rises gradually and continuously through the rest of the heliosphere out to the outer domain boundary, eventually exceeding but remaining on the order of unity. Thus, our simulation is performed in a solar/heliospheric regime that is strongly field-dominated in the inner corona, but is plasma-dominated both at the null point and far out in the heliosphere. This contrasts with previous 2.5D simulations of breakout configurations with a solar wind \citep{vanderHolst_al07, Zuccarello_al08, Zuccarello_al09, Zuccarello_al12, Soenen_al09}, in which the background and active-region field strengths were well below 10 G and the minimum plasma beta was $10^{-1}$ or larger.

 \subsection{Boundary forcing}
\label{forcing}

The system is driven by photospheric motions applied to the magnetic flux confined below the fan's south lobe. Two photospheric flows are prescribed to displace field lines along the $\phi$ coordinate in opposite directions on either side of the polarity inversion line (PIL) of the active region, thus forming a sheared arcade below the coronal null point. Each flow is defined by a cosine profile that avoids the formation of strong shearing gradients at the boundaries of the flows. Since the magnetic flux distribution below the south part of the fan is not symmetric, the photospheric flows on the two sides of the PIL do not extend symmetrically in the latitudinal direction. The velocity fields are applied over the ranges $\theta \in [24.3^\circ,28.1^\circ]$ and $\theta \in [19.1^\circ,22.4^\circ]$, respectively north and south of the PIL (right panel of Figure \ref{f-topology}). This photospheric forcing starts at $t=2.1\times 10^5~\rm{s}$, after the relaxation phase. Hereafter, we consider the starting time of this forcing as the initial time of the simulation, $t^\prime=0$. The velocity fields are applied gradually by multiplying the spatial velocity field with a time-dependent cosine ramping function. The speed of the prescribed flows on both sides of the inversion line reaches a constant value after a time interval $t^\prime_{\rm{phot}} = 1\times10^4~\rm{s}$, with a maximum driving velocity $u_{\rm{phot}}^{\rm{max}}=20~\rm{km~s}^{-1}$. This speed is greater than that of observed photospheric motions, for numerical convenience and efficiency. Nevertheless, the flows are still subsonic and highly sub-Alfv\'enic, with acoustic and Alfv\'en Mach numbers of order $10^{-1}$ and $10^{-2}$, respectively. We note that these values are close to those employed by \citet{vanderHolst_al09} in a 3D simulation of a slow streamer-blowout eruption; to obtain a fast CME, however, they had to impose footpoint flows that were faster by a factor of four.

The imposed photospheric motions inject free magnetic energy and magnetic stress into a force-balanced system. Electric currents are expected to develop along the various topological features, such as the null points and separatrices \citep{Low87,Aly90,Lau93}, and within the sheared arcade \citep[and references therein]{Forbes_al06}. These structures are described below.

\section{Formation of volume and sheet currents}
\label{current}

The left column of Figure \ref{f-nullpoint reco} displays a 2D cut in the $(r,\theta)$ plane of the $\phi$ component of the current density at  times $t^\prime =2.32$, $2.58$, and $2.74 \times 10^{4}$~s after the shearing motion has reached constant velocity. Black and white correspond to the maximum values in the positive and negative $\phi$ directions, respectively. The white structure shows the current density developing within the sheared arcade. As the simulation evolves, the shearing motion induces the elongation and strengthening of the current structure, which occurred in earlier simulations of the 2.5D breakout model \citep{MacNeice_al04,Karpen_al12}. This distributed volume current is associated with the growing inward tension force exerted by the poloidal magnetic field $\left( B_r,B_\theta \right)$, which balances the outward pressure force exerted by the shear-induced toroidal magnetic field component $B_\phi$.

The applied shearing motion also leads to the formation of a current sheet at the null point \citep{Antiochos96}, which is the thin, quasi-horizontal black structure well above the white volume current within the sheared arcade (left column of Figure \ref{f-nullpoint reco}). The right column of Figure \ref{f-nullpoint reco} shows the evolution of the separatrices (dark blue field lines) and other magnetic-field lines involved in the dynamics of the simulation (other colors, see below in \S\ref{nullpoint reco}). The shearing of the arcade causes its field lines to bulge outward, compressing and deforming the separatrices and the null point \citep{PonBatGal07a}. This leads to the misalignment of the inner and outer spines as displayed by the dark blue field lines of Figure \ref{f-nullpoint reco} \citep{GalsgaardNordlund94, Antiochos_al02, Masson_al09b}. This shearing of the spines elongates and intensifies the null-point current sheet \citep{RickardTitov96, Galsgaard_al03}. Magnetic reconnection is therefore expected to ensue first at the distorted null point.

\section{Reconnection at the null point}
\label{nullpoint reco}

The right column of Figure \ref{f-nullpoint reco} (see also the movie available in the online version) presents the connectivity evolution of selected magnetic field lines, which are plotted from fixed $r$ and $\theta$ footpoint locations at each time. Gray lines represent magnetic field whose connectivity does not change during the simulation, while the other colors represent specific connectivity domains. The red and orange field lines are initially closed below the fan's south lobe; their fixed footpoints are anchored in the negative (southern) polarity of the dipole ring. The yellow, pink, and green field lines are rooted in the negative polarity of the northern hemisphere and, respectively, are connected to the southern positive hemisphere, initially are opened to the interplanetary medium, and define the open flux belonging to the northern coronal hole throughout the evolution. 

 As the simulation evolves, magnetic reconnection takes place at the null point and the connectivity of the orange and yellow field lines changes. At $t^\prime = 2.32\times 10^4~\rm{s}$, the outermost orange line reconnects at the null point with a yellow helmet-streamer field line, which initially confines the closed outer spine. Subsequently, additional orange lines reconnect, transferring part of the flux overlying the sheared arcade (represented by the red lines) toward closed magnetic domains located on both sides of the fan's south lobe (panels d \& e of Figure \ref{f-nullpoint reco}). The yellow lines jump from the outer connectivity domain to the northern-most inner domain and form new loops below the fan's north lobe, while the orange lines pile up above the arcade connecting the two hemispheres, under the outer spine.

In the right column of Figure \ref{f-nullpoint reco}, the red and orange field lines belong to the same connectivity domain, but they reconnect with two distinct magnetic flux systems. After the orange lines have entirely reconnected with yellow lines at the null point, the red lines start to reconnect there with the pink lines. Thus, the red field lines jump from the central inner connectivity domain to the outer connectivity domain. These sheared-arcade field lines, initially closed below the fan, now open into the heliosphere; meanwhile, the initially open pink lines close down, below the fan's north lobe (panel f of Figure \ref{f-nullpoint reco}). This exchange of connectivity between open and closed field lines corresponds to the interchange reconnection mode.  
%

During the transition from the classical closed/closed (orange/yellow) null-point reconnection to the interchange closed/open (red/pink) reconnection, the magnetic topology changes. During the closed/closed reconnection, the outer spine remains closed below the helmet streamer (panel e of Figure \ref{f-nullpoint reco}), while during the closed/open interchange reconnection the outer spine has been opened to interplanetary space (panel f of Figure \ref{f-nullpoint reco}). When the entire closed magnetic flux (yellow) initially confining the null point has reconnected and been transferred to the closed domain of the fan's north lobe, the outer spine separatrix field line is positioned at the boundary between the closed (red flux) and open (pink flux) connectivity domains. The outer spine is now opened in the interplanetary medium. This topological change implies that the null-point topology is now embedded in the northern coronal hole. Thus, the sheared arcade now is positioned to erupt into an open-field region, rather than into a closed-connectivity domain as usually studied.

This interchange reconnection plays two crucial roles. First, it contributes to the flux removal, participating in the loss of equilibrium of the flux rope that is required in the breakout model of eruptions. Second, when the flux rope does erupt, its closed coronal flux is driven to interact and reconnect with the open interplanetary magnetic field.

\section{Flare reconnection and flux-rope eruption}
\label{fr eruption}

As the null-point reconnection progresses, the volume current within the sheared arcade becomes more distended, thinner, and more intense, due to the ongoing photospheric shearing motion (\S\ref{current}). This evolution is evident in Figure \ref{f-nullpoint reco} and accelerates strongly at the later times shown in Figure \ref{f-fr eruption}. When the volume current becomes very distended, its central section thins to a quasi-vertical current sheet that gets squeezed to the grid scale. The sheared arcade then reconnects internally, signaling the onset of the flare-reconnection phase \citep{Forbes00,Karpen_al12}. As in the schematic CSHKP model for eruptive flares (see \S\ref{intro} and Figure \ref{f-cartoon}), two new flux systems are born: the post-flare loops, forming an arcade that closes down below the flare-reconnection site; and twisted field lines high in the corona, forming the ejected flux rope. Due to the azimuthal symmetry of the 2.5D simulation, the new flux-rope field lines are disconnected completely from the photosphere.

The colored field lines plotted in Figure \ref{f-fr eruption} are identical to those in Figure \ref{f-nullpoint reco}. The field lines are over-plotted on a 2D cut in the plane ($r,\theta$) of the $\phi$ component of the current density, $j_\phi$, shaded in grayscale. The evolution of $j_\phi$ shows the evolution of the flux rope, while the colored field lines show the dynamics of the magnetic flux systems involved in the eruption. The six panels of Figure \ref{f-fr eruption} display different times of the simulation -- $t^\prime = 3.11$, $3.16$, $3.17$, $3.18$, $3.20$, and $3.26 \times 10^4$~s, respectively -- starting after the opening of the outer spine. (A movie is available in the electronic version of this article.) 

At $t^\prime = 3.11\times10^4~\rm{s}$, the $j_\phi$ structures are similar to those observed during the null-point reconnection (see Figure \ref{f-nullpoint reco}), except that the apex of the volume-current  structure has acquired an  ellipsoidal shape. The imminent formation of the disconnected flux rope suggests that the flare reconnection may have started already at this time. Examining our MHD simulation data at a high temporal cadence of 100 s, we looked for a magnetic O-point in the numerical domain that would constitute the central axis of the flux rope. At time $t^\prime = 3.07\times10^4~\rm{s}$, an O-point appeared at the top of the flare current sheet, confirming that the internal reconnection of the sheared arcade indeed had begun and that the flux rope was beginning to form.

Between $t^\prime = 3.11\times 10^4~\rm{s}$ (Figure \ref{f-fr eruption}a) and $t^\prime = 3.18\times 10^4~\rm{s}$ (Figure \ref{f-fr eruption}d), the ellipsoidal current structure enlarges as the flare current sheet thins. This implies that sheared-arcade field lines are reconnecting at the flare current sheet. The newly reconnected field lines wrap around those that reconnected previously,  increasing the flux-rope magnetic flux. Meanwhile, the flux transfer enabled by the null-point and interchange reconnections (\S\ref{nullpoint reco}) induces a decrease in the downward magnetic tension of the overlying  magnetic field. These changes in the force balance drive the rise and eventual take-off of the flux rope \citep{Antiochos_al99, MacNeice_al04,Karpen_al12}.

\begin{figure}
\centerline{
 \includegraphics[width=0.4\textwidth,bb=133 58 494 778,clip=]{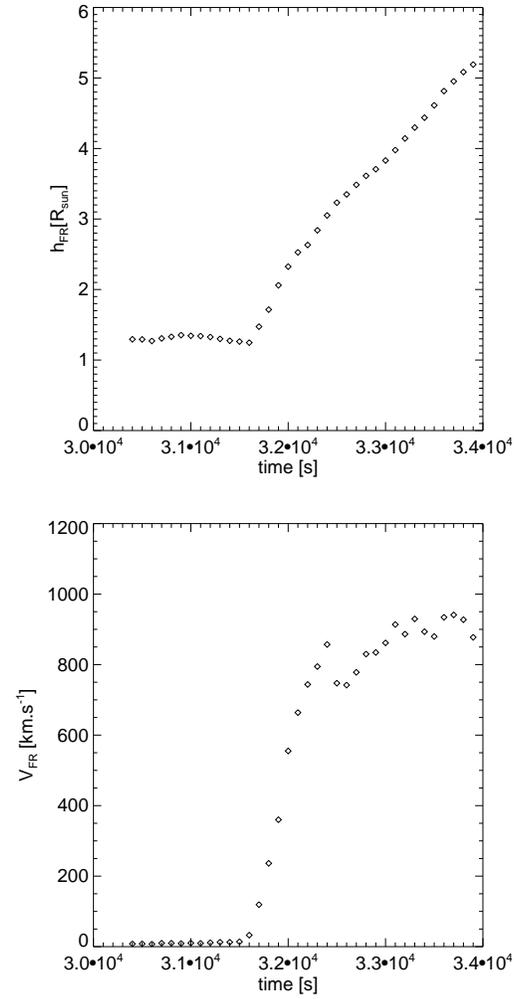}
 }
\caption{Flux-rope trajectory. The top panel displays the temporal evolution of the height of the O-point of the flux rope; the bottom panel displays its speed.
}
  \label{f-speed}
\end{figure}

In Figure \ref{f-fr eruption}, the time interval between panels a and b is $500~\rm{s}$, while that between panels b, c, and d is only $100~\rm{s}$. Thus, the evolution of the flux-rope current structure exhibits a very distinct transition during this early development (see also the movie in the electronic version). Prior to time $t^\prime = 3.16\times 10^4~\rm{s}$, the flux rope rises in the corona only very slowly. At time $t^\prime = 3.17\times 10^4~\rm{s}$ and later, the flux rope has strongly accelerated and is rising very fast.
 We evaluated the radial location and flow speed of the plasma at the location of the O-point during this interval, shown in Figure \ref{f-speed}. Between $t^\prime = 3.07\times 10^4~\rm{s}$ (initial appearance of the O-point) and $t^\prime = 3.16\times 10^4~\rm{s}$, the flux rope does not move significantly,  remaining roughly at $h_{FR}= 1.2R_\odot$ with a speed of $V_{FR}\simeq 20~\rm{km~s}^{-1}$. Between $t^\prime = 3.16\times 10^4~\rm{s}$ and $t^\prime = 3.24\times 10^4~\rm{s}$, the flux rope rapidly ascends in the corona as its velocity increases to $V_{FR}(3.2R_\odot)\simeq 900~\rm{km~s}^{-1}$ in $800~\rm{s}$.  The average flux-rope acceleration during this interval is $a_{FR}= 1.1~\rm{km~s}^{-2}$. Thereafter, the CME moves out at a roughly constant speed of $V_{FR}\simeq 900~\rm{km~s}^{-1}$ beyond $h_{FR}= 3.2R_\odot$. Our simulation clearly yields a strongly accelerated, fast eruption.

The evolution of our erupting flux rope is consistent with the three distinct phases described for observed CMEs \citep{Zhang_al01}: the initiation phase, when the flux rope slowly rises in the corona; the rapid acceleration of the CME in the corona, corresponding to the impulsive phase; and finally the propagation of the CME in the interplanetary medium at a constant speed, the terminal phase.

The fast eruption induces the formation of a new current sheet at the front of the flux rope (panel c in Figure \ref{f-fr eruption}). This current sheet results from the formation of a compression region between the front of the erupting flux rope and the ambient medium traveling at the speed of the isothermal solar wind. At the time of eruption onset, the plasma speed at the leading edge of the flux rope is $V_{LE} (\simeq1.76R_\odot)\simeq 150~\rm{km~s}^{-1}$, significantly faster than the local solar wind speed, $V_{SW}\simeq 100~\rm{km~s}^{-1}$. Thus, the magnetic field is expected to be squeezed in this region, consistent with the development of a current sheet. However, comparing the flux-rope velocity with the local Alfv\'en speed, $c_A\simeq 2000~\rm{km~s}^{-1}$, we conclude that this compression region is not a shock.

\section{Interaction between the CME flux rope and the interplanetary medium}
\label{multi reco}

The right column of Figure \ref{f-fr eruption} shows the dynamics of the flux rope that propagates out of the corona and into the interplanetary medium after the eruption. The field lines representing the open flux of the northern polar coronal hole in Figure \ref{f-fr eruption} are colored as a function of their $B_\phi$ component, with green and blue corresponding to $|B_\phi| = 0~\rm{G}$ and  $|B_\phi| = 0.02~\rm{G}$, respectively. The initial magnetic field (Figure \ref{f-topology}) is strictly poloidal, lying in the $\left( r, \theta \right)$ plane and having $B_\phi \equiv 0$. The photospheric footpoint motions introduce a nonzero $B_\phi$ into the system only within the central sheared arcade, below the south lobe of the fan. Part of this toroidal flux resides on the field lines that reconnect at the flare current sheet and become entrained into the disconnected flux rope (\S \ref{fr eruption}). If the ejected flux rope subsequently reconnects with an open line of the background field, that initially green line will acquire a nonzero $B_\phi$ component and become partially blue.

In panel d of Figure \ref{f-fr eruption}, the innermost open green field line becomes blue all along a section that has a helical shape. The blue/green color and the morphology show that this previously open line has reconnected with the flux rope. The newly reconnected field line belongs both to the flux rope and to the coronal hole, coupling the flux rope to both the base of the corona within the polar hole and the remote heliosphere at the far end of the open field line. As the flux rope propagates through the northern coronal hole, this reconnection continues to process open flux, forming additional new field lines that connect the flux rope to the solar pole and to the interplanetary medium (panels e and f in Figure \ref{f-fr eruption}). This third episode of reconnection, occurring between the flux rope and the open field, is similar to the interchange reconnection in a standard closed null-point topology. Since it involves closed CME field lines and open coronal-hole flux, we refer to it as the CME-interchange reconnection mode.

Meanwhile, a fourth reconnection episode occurs. The field lines closed below the fan's north lobe (pink lines) and the open field to the south of the outer spine line (red lines) reconnect together. This reconnection is of the classical null-point interchange type, and the two resulting reconnected fluxes are the closed red arcade lines and the open pink flux shown in panels e and f of Figure \ref{f-fr eruption}. This episode is a direct consequence of the eruption of the flux rope and is due to the re-formation of the low-coronal null point (see \S\ref{evolution topology} for details). After the red and pink fluxes have been processed fully in this way, reconnection across the re-formed null point continues between the yellow and orange flux systems. The latter reconnection is of the classical closed/closed type, and completes the restoration of the helmet streamer configuration (Figure \ref{fr eruption}f) from which the event began (Figure \ref{f-nullpoint reco}).

At the end of the simulation, the reconnection between the flux rope and the open field terminates well before all of the magnetic flux in the coronal hole has reconnected. The magnetic flux of the flux rope is much smaller ($\Phi_{FR} \simeq 1.96 \times 10^{20}~\rm{Mx}$) than the open flux ($\Phi_{open} \simeq 2.26 \times 10^{22}~\rm{Mx}$). Therefore, the flux rope entirely reconnects and disappears as a separate entity, while a great deal of unsheared open flux remains in the northern coronal hole. The total merger of a flux rope with the open interplanetary field through reconnection may occur during the eruption of faint CMEs with small magnetic flux. However, we do not believe that it should be the conventional evolution for all CMEs. For our simulation, we expect that the twist added to the open field lines will be dispersed through torsional Alfv\'en waves well before the magnetic structure reaches the Earth. This would be in disagreement with the detection at Earth of interplanetary coronal mass ejections/magnetic clouds that are the counterpart of the CMEs launched during solar eruptions \citep{Wimmer_al06}.  

\section{Evolution of the magnetic topology}
  \label{evolution topology}
 
 The reconnection dynamics described above give rise to a sequence of topological transitions in the magnetic field. We note that throughout this paper, the words ``fan'' and ``spine'' have been used to denote the different separatrix structures of the magnetic field. It is important to keep in mind that these terms actually are defined for a 3D null-point topology, rather that the 2.5D null-line geometry of our simulation. In 3D, the fan lines form a surface while the spines are singular lines; in 2.5D, on the other hand, both structures are surfaces. Thus, there is no real physical distinction between the ``fans'' and ``spines'' in the simulation presented in this paper. We still employ those terms, however, in order to develop insight into the generalization of our results to fully 3D configurations.

 Figure \ref{f-new topo} displays the six topologies that successively develop during the evolution. As before, separatrix field lines are plotted in dark blue, while other field lines are color-coded as in Figures \ref{f-nullpoint reco} and \ref{f-fr eruption} to show the same flux domains. Light blue field lines are added to Figure \ref{f-new topo} to represent the sheared flux of the inner arcade and the twisted CME flux rope. The initial magnetic field corresponds to a closed null-point topology, associated with the null point NP1 and its closed outer spine line (Figure \ref{f-new topo}a). The classical null-point separatrices delimit the different connectivity domains (see \S\ref{topology}). As described in \S\ref{fr eruption}, the subsequent null-point reconnection at NP1 eventually opens the outer spine to the heliosphere, transforming its topology into the open null-point type (Figure \ref{f-new topo}b).

Following the onset of the eruption, the CME flux rope forms through flare reconnection, implying the pairwise formation of an O-point O1 and a new X-type null point NP2 (Figure \ref{f-new topo}c). O1 is located at the central axis of the flux rope, and NP2 is created in the flare current sheet below the flux rope, where the magnetic field lines of the sheared arcade are anti-parallel. Accordingly, new separatrix surfaces associated with O1 and NP2 appear in the system and define new connectivity domains. The flux rope belongs now to an independent flux domain, delimited by a new separatrix S1 in Figure \ref{f-new topo}c. The post flare-reconnection loops below the erupting rope are delimited from the highly stressed original arcade by the separatrix S2, which is joined to S1 by the current sheet containing NP2 (Figure \ref{f-new topo}c). 

 Although the two X-type null points, NP1 and NP2, share some properties, we emphasize that their physical origin and topological roles are very different. NP1 is a robust topological feature due to the multipolarity of the field, and must be present independent of whether the system is ideal or dissipative. The breakout current sheet forms as a consequence of the deformation of NP1. NP2, on the other hand, is not a robust topological feature. First, the flare current sheet forms as a result of the extreme stretching of the inner arcade field lines; then, NP2 and O1, and almost certainly many other null points, form inside this current sheet as a result of dissipation. NP2 and O1 would not occur under a truly ideal evolution and, in principle, they could disappear by simply merging with each other. In contrast, NP1 can never disappear, even in 3D. Note also that the flare current sheet associated with NP2, just like the breakout current sheet associated with NP1, has much more structure than is illustrated in the figures. During the highly dynamic phase of the simulation, both current sheets break up into multiple magnetic islands, each of which defines its own separatrix. These islands, however, eventually merge with the flux at either end of the current sheets. Consequently, we show in Figure \ref{f-new topo} only the time-averaged, global separatrix structure.

 During the early rising phase of the flux rope (\S\ref{fr eruption}), the current sheet due to the deformed null point NP1 becomes increasingly elongated, which increases the rate of breakout reconnection. Note that this reconnection acts to decrease the flux inside S1, {\it i.e.}, in the erupting flux rope. On the other hand, the rate of flare reconnection at NP2 increases even more rapidly, which results in a net increase in the flux inside S1. This competition between the breakout and flare reconnection continues until the two separatrices that can be seen at the bottom center of Figure \ref{f-new topo}c -- the inner spine surface on the left and the fan surface on the right -- converge and merge at the flare current sheet NP2. 

The magnetic topology at this instant of merger is shown in Figure \ref{f-new topo}d. Note that the topology is degenerate, in that all of the separatrices and the two null points are connected. This instant also corresponds to the disconnection of the CME flux rope from the flare loop arcade, at least in 2.5D. Consequently, the flux in the ejecta stops growing at this time; from now on, that flux only decreases, via CME-interchange reconnection at NP1.  Note also that the null points NP1 and NP2 switch nature at this time. NP2 now becomes the robust null point associated with the multipolarity, while NP1 becomes the transitory null point that can disappear by merging with O1 (Figure \ref{f-new topo}d). In fact, this is exactly what occurs later in the simulation.
 

The start of the fourth reconnection episode (\S\ref{multi reco}) disrupts the merged separatrices of Figure \ref{f-new topo}d and forms a new topology. We use the label $\Sigma$ for the new separatrices. The exchange of connectivity between pink and red field lines, at the null point NP2, transfers the open outer spine from the left (north) side of the flux rope to the right (south) side (Figure \ref{f-new topo}e). Thus, the open null-point topology previously associated with the original null point NP1 (Figure \ref{f-new topo}b) now belongs to the flare-sheet null point NP2 and its inherited inner and outer spines (Figure \ref{f-new topo}e). The original null point NP1 now lies on the separatrix surface $\Sigma$1 separating the disconnected CME flux rope from the background open flux. NP1 acquires a spine that splits from the fan below NP1 and the open outer spine above NP1 (Figure \ref{f-new topo}e). 

The CME-interchange reconnection episode (\S\ref{multi reco}), which couples the flux rope to the open magnetic field, does not change the new topology. This reconnection occurs at the null point NP1 and transfers open magnetic flux through the $\Sigma$1 spine and separatrices to the far (south) side of the flux rope (Figure \ref{f-new topo}f). As indicated in the figure, although the reconnection does not alter the topology, it does transfer sheared light blue field lines from the disconnected flux rope to the open heliospheric magnetic field.

%


\section{Discussion}
\label{discussion}

To understand how flare-accelerated particles can directly access the open interplanetary magnetic field during a solar eruption producing a CME, we investigated the dynamics of the magnetic reconnection when a flux rope forms, erupts, and propagates into the solar wind. Our aim is to determine how particle injection channels are opened from the flare site to the remote heliosphere in response to the large-scale dynamics of the eruption. We do not consider the details of the flare acceleration mechanism or of the population of the resulting energetic particles; we simply assume that the SEPs are accelerated at the flare-reconnection site, and ask how such particles can promptly access the open IMF in the aftermath of an eruption. Our axisymmetric 2.5D MHD simulation in spherical coordinates is based on a null-point topology susceptible to breakout-CME initiation, embedded below a helmet streamer formed by a simple isothermal solar wind.

We have analyzed the detailed topological changes associated with a fast breakout CME erupting into an interplanetary magnetic field opened by the solar wind. The initial magnetic field is a null-point topology with a closed outer spine that is confined below the helmet streamer.  Prior to the initiation of the CME, the magnetic energy builds up due to slow footpoint shearing motions imposed within the arcade below the null point. The excess energy so introduced inflates the sheared arcade, distorting the null point above it into an extended current sheet and eventually inducing magnetic reconnection there. After all of the closed flux below the streamer top and above the null point has reconnected, the initially closed outer spine opens into the heliosphere: the null-point topology now is surrounded by open field in the northern coronal hole. Additional null-point reconnection occurs gradually, until the inner sheared arcade becomes highly distended and narrows at high altitudes to develop an internal (flare) current sheet.  The onset of reconnection across this sheet causes the system to lose its equilibrium, rapidly forming a flux rope that erupts into the corona at high speed and propagates away into the interplanetary medium. During this evolution, additional episodes of reconnection occur that reform the open-spine null-point topology at low altitudes and, most important, couple the closed field lines within the CME flux rope to the open field lines of the interplanetary medium.

The fast nature of our CME is noteworthy as an exception to the results of previous 2.5D simulations of breakout CME eruptions into the solar wind \citep{vanderHolst_al07, Zuccarello_al08, Zuccarello_al09, Zuccarello_al12, Soenen_al09}. The major difference between those studies and ours is the strength of the assumed background and active-region magnetic fields: we assumed field strengths that are larger by about a factor of 10, with a concomitant difference by a factor of 100 in the plasma beta. Thus, it is not surprising that our strongly magnetically dominated corona yields a much more explosive, higher-speed eruption. 
The slow streamer-blowout simulation performed by \cite{Zuccarello_al12} nevertheless exhibits magnetic-reconnection dynamics similar to those discussed in this paper. After the eruption of their CME, which results from the breakout and flare reconnection, the flux rope reconnects with the closed magnetic field below the helmet streamer. 
The evolution described in \cite{Zuccarello_al12} follows the same topological changes as ours described in \S\ref{evolution topology}, except that their flux-rope separatrix surface S1 remains closed within the helmet streamer, rather than reconnecting out into the wind-opened corona.
The results obtained with 3D MHD simulations \citep{vanderHolst_al09, Cohen_al10, Lugaz_al11} should display an analogous sequence of changes as the CME propagates into the interplanetary medium. However, the evolution of the magnetic topology has not been clearly established in those studies. The complexity of the three-dimensional topological objects makes the determination of the associated changes a far more challenging task than in the much simpler 2.5D geometry assumed in this paper.
  
In the standard CSHKP model of eruptive flares, the flare-accelerated particles are trapped in the CME and do not have access to the open interplanetary medium (see \S\ref{intro}). However, this standard model does not consider the CME's interaction with and propagation into the ambient magnetic field. Our study demonstrates that the dynamics of magnetic reconnection, by inducing a specific topological evolution, can couple the magnetic field of a CME to the open interplanetary magnetic field. The resulting reconnected field lines are rooted at the solar surface within the pre-existing coronal hole, connect to the remote heliosphere, and pass through the CME flux rope, providing a path for the prompt escape of flare-accelerated particles to the Earth. 

Energetic particles in the corona have a small Larmor radius and follow the reconnected magnetic field lines. Electromagnetic emissions (UV, X-ray, $\gamma$-ray) occurring when particles impact the denser chromospheric layer are localized at the footpoints of the reconnected field lines \citep[see e. g.][]{Demoulin_al97,Masson_al09a,Reid_al12}. Therefore, it is well established that the reconnecting magnetic field effectively channels the particles energized by flares. In the magnetic configuration of our MHD simulation, the particles initially trapped in the flux rope field lines can access the interplanetary medium through the newly reconnected field lines coupling the open and closed fields. Therefore, energetic particles traveling initially along the CME flux-rope field lines may be injected onto the newly reconnected field lines that open to the interplanetary medium.

Energetic particles in the corona generate radio emissions. The coronal localization of these radio sources, combined with complementary observations, provides observational diagnostics on the location of energetic electrons and their related solar phenomena \citep{PickVilmer08}. Recently, \cite{Demoulin_al12} performed a detailed analysis combining radio, white-light, hard X-ray, and EUV observations of the CME on 2001 April 15. In this study, they highlighted specific radio emissions that correspond to the flare reconnection occurring below the flux rope. They also found that the CME flux rope observed in white light was simultaneously observed at radio wavelengths; radio CMEs were discovered by \cite{Bastian_al01} and confirmed by \cite{Maia_al07}. \cite{Demoulin_al12} argued that the radio emission of the CME flux rope is produced by energetic electrons, initially accelerated by the flare reconnection and injected along reconnected field lines wrapping around the flux rope. Finally, they showed that several radio sources appear during the post-eruptive phase, on the edges of the radio CME. They argued that these radio sources correspond to energetic electrons accelerated during magnetic-reconnection processes between the CME and the ambient magnetic field. In addition, these multiple episodes were temporally consistent with the injection of energetic electron beams into the interplanetary medium derived from type-III radio-burst observations \citep{PickVilmer08}.

Our numerical results on the reconnection dynamics of a CME propagating into the interplanetary medium are consistent with the observational results and interpretation of \cite{Demoulin_al12} concerning both the flare reconnection (\S\ref{fr eruption}) and the development of multiple reconnection sites between the CME and the ambient field (\S\ref{multi reco}). It is noteworthy that new injections of energetic electrons into the interplanetary medium occurred almost simultaneously with the inferred multiple reconnection episodes. This evidence strongly supports the new model for particle injection established in this paper, in which the coupling between the closed CME flux-rope magnetic field and the open field of the solar wind provides a path for energetic particles to escape.

Our model magnetically reconfigures the corona through multiple reconnection episodes, providing theoretical support for a process already proposed to explain observations during solar energetic particle events.  \cite{Klein_al11} showed that western CME-less flares, accelerating particles in the corona, do not produce any SEP events at the Earth. Thus, CMEs may be an essential ingredient for particle escape into the interplanetary medium. Those authors argued that, among other mechanisms, magnetic reconnection between the closed CME and the open interplanetary magnetic field may be required to inject particles. During CME-less flares, this coupling cannot occur, leading to particle confinement in the corona and no SEPs. 

During some SEP events, energetic particles can be detected farther than $90\degr$ from the flaring active region \citep{Lin70,Kallenrode_al92,Nitta_al06,Rouillard_al11}. Usually, these in-situ measurements are ascribed to CME-driven shock acceleration that also injects energetic particles over a wide longitudinal range. The scenario described in this paper provides a new perspective on these observations. Indeed, the coupling between open and closed fields does not occur near the active region in our simulation, but far away from it. Rather than injecting energetic particles through a CME-driven shock over a wide longitudinal range, multiple reconnections between the CME (where flare-accelerated particles are trapped) and open magnetic field can allow the particles to escape from locations remote from the original flare site. This would be especially true if the open field had a corridor topology as described in \citet{Antiochos_al11}, which can map to a large longitudinal arc in the heliosphere as suggested by several studies of SEPs \citep{Pick_al06,Klein_al08,Dresing_al12,Wiedenbeck_al13}.


 In addition to these spatial properties, we can constrain the temporal evolution of the particle escape. Assuming that the particles are accelerated at the beginning of the impulsive phase of the eruption, $t=3.16\times10^4\rm{s}$, and injected at the beginning of the CME-interchange reconnection episode, $t=3.173\times10^4\rm{s}$, the time interval between the acceleration and the injection is of order $\Delta t_{simu}= 130~\rm{s}$. This time interval corresponds to the time required to process all the closed flux overlying the flux rope, after the onset of the impulsive phase. The closed overlying magnetic flux, $\Phi=\overline{B}~d$, is processed at $\simeq 400~\rm{km.s}^{-1}$, which is $ \simeq 10\%$ of the local Alfv\`en speed at the vicinity of the flare-reconnection site. The distance between the flare-reconnection site, located at $r= 1.224~\rm{R_\odot}$, $\theta_2 = 0.442~\rm{rad}$  and the last closed flux, $\theta_1 = 0.3892~\rm{rad}$, is $d = 45~\rm{Mm}$. Thus, one can estimate the reconnection time, $t_{reco} = \overline{B}~d / (0.1~C_A~\overline{B}) = 112.5~\rm{s}$.  The result obtained for $\Delta t_{simu}$ is consistent with $t_{reco}$, therefore, we can rescale the eruption to an observed event based on the Alfv\'en characteristic time scale as in  \citet{Karpen_al12}. The active region in the simulation latitudinally extends over $\simeq 20^\circ$, {\it i.e.} the typical length $L_{simu} \simeq 240\times 10^6~\rm{m}$, whereas the size of an observed active region is of order $100\times10^{6}~\rm{m}$. The Alfv\`en speed in the active region is on average $\simeq 1000~\rm{km.s}^{-1}$, which is in the lower range of the Alfv\`en speed estimated in  observed active regions. Therefore, the timescale for an observed eruption,  $t_{obs}= L_{obs}/C_{A,obs}$, is smaller by a factor 2.4, or more, than the time-scale of the simulation $t_{simu} =L_{simu}/ C_{A,simu}$. Note that larger magnetic field strength or a smaller active region will decrease the scaling factor, leading to temporal evolution of the simulation closer to the observations. Applying this scaling factor, the time interval between the acceleration and the injection is of $\Delta t_{obs}= 130/2.4~\rm{s} \simeq 54~\rm{s}$ for an observed event. This suggests that the flare-accelerated particles can be  promptly injected into the heliosphere, which is consistent with observational studies of SEPs \citep{Kahler_al01,Klein_al05,Masson_al09a}. However, it should be stated that the simple analysis above neglects several factors that may be important in actual SEP events. First, the relative timing and location between the flare and CME-interchange reconnection episode can be different for a 3D geometry. Second, particles can be accelerated later during the impulsive phase and, therefore, are not necessarily injected at the beginning of the CME-interchange.

The primary conclusion from our study is that, in the presence of nearby open flux, interchange-type reconnection between an erupting flux rope and the open flux naturally leads to the escape of flare accelerated particles. Our results, however, raise major theoretical and observational issues. On the theoretical side, the most important issue is the extension of the model to 3D. A key shortcoming of our 2.5D simulation is that the erupting rope disappears completely as a result of reconnection with the open flux, Figure~\ref{f-fr eruption}; whereas, ICMEs/magnetic clouds are observed in association with impulsive SEPs at 1~AU. However, in 3D the breakout and the interchange reconnection will occur at the current sheet formed from a true null point and is likely to show strong variation along the axis of the erupting flux rope, which itself will be fully 3D. We conjecture that in this case interchange reconnection will not follow the one-to-one ordered reconnection imposed by the 2D geometry, but may well occur with some flux from deeper inside the flux rope, forming a complex ejecta entangling open and closed flux. This conjecture will be tested with future fully 3D simulations.


On the observational side, the most important issue raised by our model is the quantitative distribution of flux required for flare-particle escape. Note that in order for the erupting flux rope to undergo interchange reconnection with open field, the overlying closed flux has to be removed. In the pre-eruption configuration shown in Figure~\ref{f-topology}, the amount of overlying closed flux is small compared to the amount of erupting flux; in other words, the coronal hole is ``near'' the eruption and, hence, interchange reconnection easily occurs. On the other hand, if the amount of overlying closed flux is large then interchange reconnection becomes highly unlikely. Therefore, comparison of our model with observed events and prediction of whether flare particles will or will not escape requires detailed knowledge of the amount of flux in the various coronal systems prior to eruption. In principle, such information can be obtained from high-resolution magnetograms, but again, fully 3D eruption models will be required, because concepts such as ``overlying'' and ``near'' become much more subtle in 3D. The 3D simulations, therefore, will provide the definitive observational tests of our model for flare particle escape.

One of us (SM) gratefully acknowledges support from the NASA Postdoctoral Program, administered by Oak Ridge Associated Universities through a contract with NASA, during her stay at NASA Goddard Space Flight Center. This work was supported, in part, by the NASA TR\&T and SR\&T Programs.

\appendix
\section{Analytical expression for the dipole ring}
\label{appendix}

In cylindrical coordinates ($\xi,\phi, z$), the vector potential for a toroidal ring of magnetic dipoles is ${\bf A}= A_\phi(\xi,z) {\bf e_\phi}$, where ${\bf e_\phi}$ is the unit vector along the ${\phi}$ axis. It is defined by the integral

\begin{eqnarray}
A_\phi(\xi,z) = B_0 d^2 R_{\odot} 
\int^{\phi+\pi}_{\phi-\pi}\frac{{\mu_z} \xi - \left[{\mu_z} a + {\mu_\xi} (z-b) \right] \cos (\chi - \phi)}{\left[ \xi^2 + a^2 + (z-b)^2 - 2a\xi \cos(\chi-\phi) \right]^{3/2}} d\chi
\label{eq-Aphi1}
\end{eqnarray}

\noindent where $B_0(\mu_\xi{\bf e_\xi}+\mu_z{\bf e_z})$ is the magnetic moment, with $B_0$ its field magnitude and ${\mu_\xi}$, ${\mu_z}$ its direction cosines respectively along the $\xi$-axis and $z$-axis, such that ${\mu_\xi}^2+{\mu_z}^2=1$; $a$, $b$, and $\chi$ are respectively the $\xi$, $z$, and $\phi$ positions of the dipole in the cylindrical coordinate system; and $d$ and $R_{\odot}$ are respectively the depth of the dipole below the surface and the solar radius. Introducing the new variable $\psi=(\chi-\phi+\pi)/2$, where $\chi-\phi$ represents the angular separation between the local contribution to the ring dipole and the location of the point where the vector potential is computed, we have

\begin{eqnarray}
\cos(\chi-\phi) = \cos(2\psi-\pi) 
= -\cos(2\psi) = -1+2\sin^2\psi.
\end{eqnarray}

\noindent With this variable change, the vector potential becomes

\begin{eqnarray}
A_\phi(\xi,z) = 2 B_0 d^2 R_{\odot} 
\int^{\pi/2}_{0}\frac{ \left[ {\mu_z} (\xi+a) + {\mu_\xi} (z-b) \right] - 2 \left[ {\mu_z} a + {\mu_\xi} (z-b) \right] \sin^2 \psi }{\left[(\xi + a)^2 + (z-b)^2 - 4a\xi \sin^2\psi\right]^{3/2}} d\psi,
\label{eq-Aphi2}
\end{eqnarray}

\noindent after exploiting the symmetry of the integrand of Eq.\ (\ref{eq-Aphi2}) to reduce the integration range from $[0,\pi]$ to $[0,\pi/2]$. Introducing the elliptic parameter $k$,

\begin{eqnarray}
k \equiv \frac{4a\xi}{(\xi+a)^2+(z-b)^2},
\end{eqnarray}

\noindent we rearranged Eq.~(\ref{eq-Aphi2}) and reduced it \citep{GradshteynRyzhik65} to complete elliptic integrals of the first and second kinds, $K(k)$ and $E(k)$,
\begin{eqnarray}
K(k) &=& \int^{\pi/2}_0 \frac{d\psi}{\sqrt{1-k \sin^2\psi}}, \nonumber \\
E(k) &=& \int^{\pi/2}_0 d\psi \sqrt{1-k \sin^2 \psi}.
\end{eqnarray}

\noindent The following final expression for the vector potential of the dipole ring was obtained:
\begin{eqnarray}
A_\phi(\xi,z) = \frac{ 2 B_0 d^2 R_{\odot}}{\left[ (\xi+a)^2+(z-b)^2\right]^{1/2}\left[(\xi-a)^2+(z-b)^2\right]} \times \phantom{------} \nonumber \\
\Bigl\{ \left[ (\xi+a) \mu_z + (z-b) \mu_\xi \right] E(k) - 2 \left[ a \mu_z + (z-b) \mu_\xi \right] \left\{ K(k) + k^{-1} \left[ E(k) - K(k) \right] \right\} \Bigr\}  
\label{eq-Aphi3}
\end{eqnarray}

\begin{figure}
\centerline{
 \includegraphics[width=1.\textwidth,clip=]{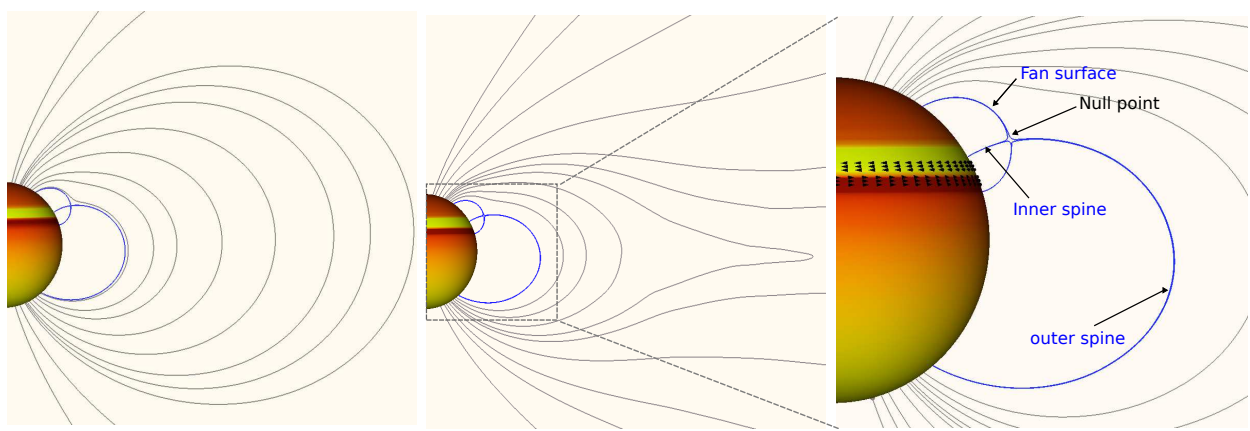}
 }
 \caption{Pre- and post-relaxation magnetic fields.  Left and middle panels: the magnetic configuration before turning on the solar wind at $t = 0$ s and after the system reaches its quasi-steady state at $t = 2.1 \times 10^5$ s, respectively. The radial magnetic field at the photospheric surface is color-shaded, with $B_r \in [-10, 20]$. Gray lines display magnetic field that initially is totally closed (left panel), but subsequently opens at the solar poles (middle panel); dark blue lines display the magnetic separatrices. Right panel: a zoomed-in view of the null-point topology. Important topological features are labeled, and the location and direction of the photospheric forcing are indicated by black arrows on the photospheric boundary.}
  \label{f-topology}
\end{figure}

\begin{figure}
\centerline{
 \includegraphics[width=0.8\textwidth,clip=]{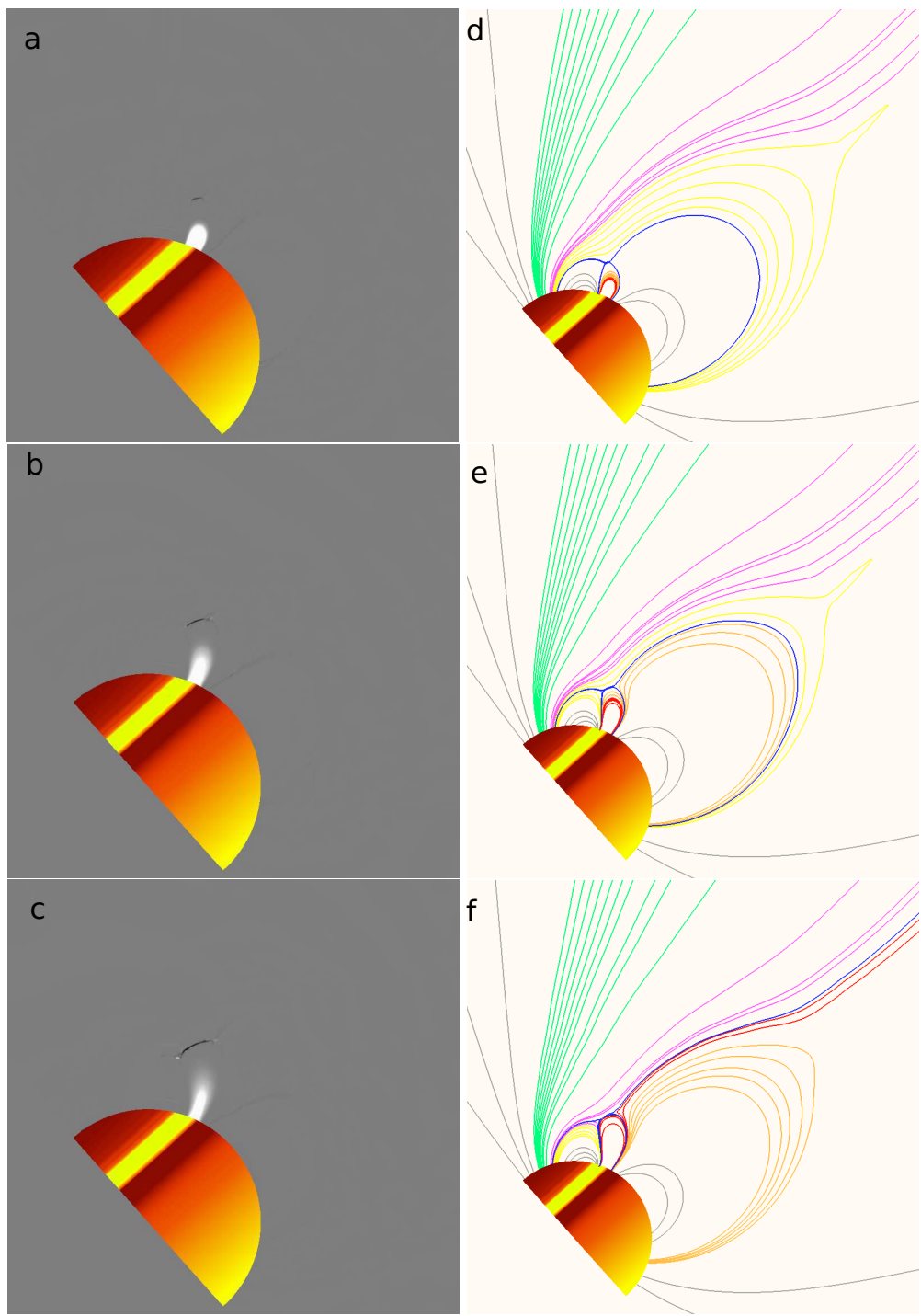}
 }
\caption{Reconnection at the null point.  The temporal evolution of the $\phi$ component of the current density (left column) and selected magnetic field lines (right column) are shown after onset of the photospheric shearing motions. The radial magnetic field at the photospheric surface is color-shaded as in Figure \ref{f-topology}. The $\phi$ component of the current density is gray-shaded; black and white correspond to $R_{\odot} {\bf \phi} \cdot \nabla \times {\bf B} = -7$ and $+7$, respectively. For the color scheme used to draw the field lines, see text. 
}
\label{f-nullpoint reco}
\end{figure}

\begin{figure}
\centerline{
 \includegraphics[width=0.8\textwidth,clip=]{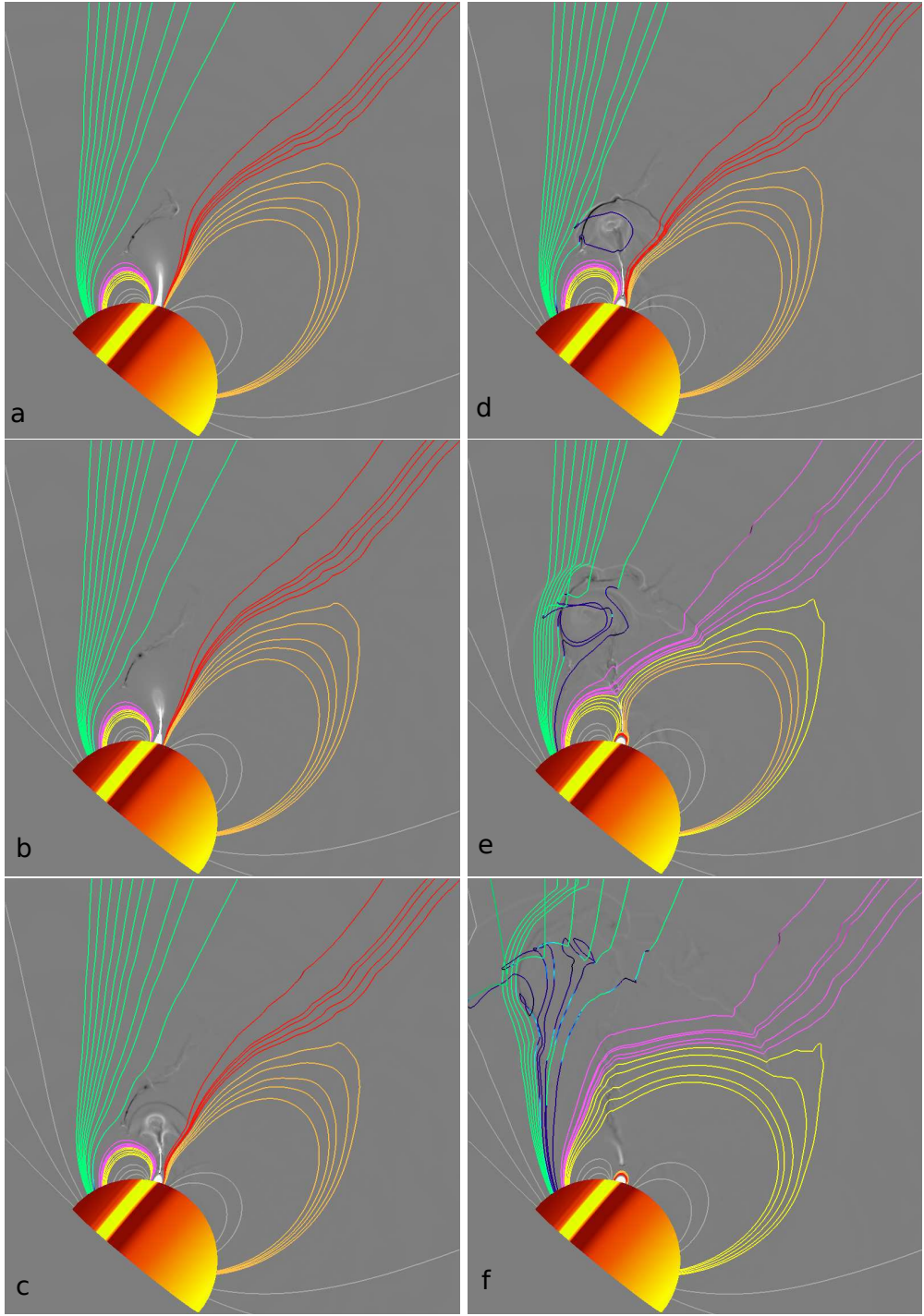}
 }
  \caption{Flare reconnection and flux-rope eruption. The radial magnetic field at the photospheric surface is color-shaded as in Figures \ref{f-topology} and \ref{f-nullpoint reco}, and the $\phi$ component of the current density is gray-shaded as in Figure \ref{f-nullpoint reco}. Field lines are drawn as in Figure \ref{f-nullpoint reco}, except that the northern coronal-hole field lines now are colored according to the value of their $\phi$ component of the magnetic field: green and blue correspond, respectively, to $\vert B_\phi \vert = 0$ G and $\vert B_\phi \vert = 0.02$ G. (An animation is available in the online journal.)}
 \label{f-fr eruption}
 \end{figure}

\begin{figure}
\centerline{
 \includegraphics[width=0.9\textwidth,bb=25 123 581 641,clip=]{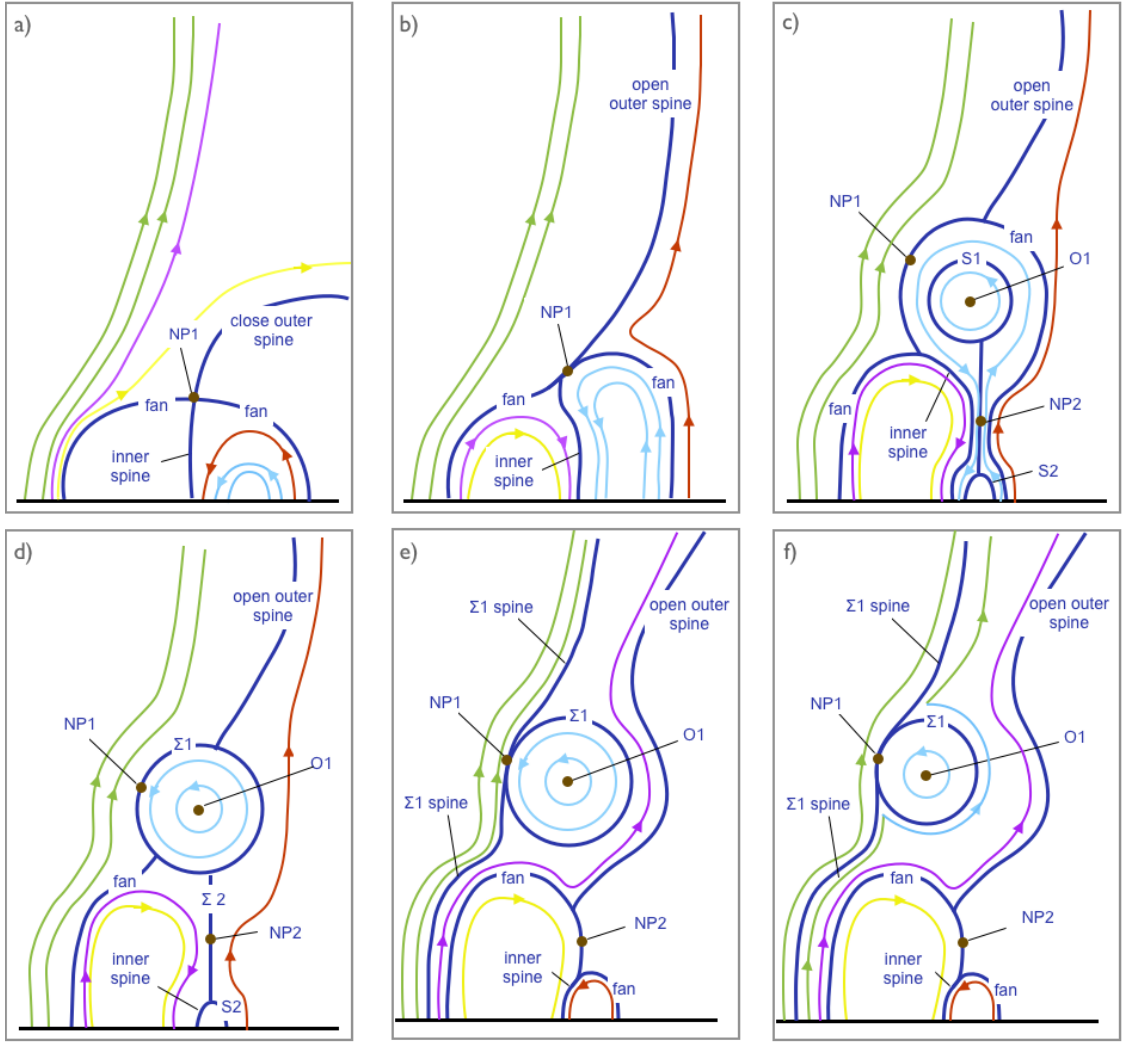}
 }
\caption{Schematic of the topology evolution during the CME eruption and coronal propagation. Light blue lines represent inner-arcade field lines with shear; the other field lines follow the same color scheme as in Figures \ref{f-nullpoint reco} and \ref{f-fr eruption}. Panels illustrate: a) the initial null-point topology with closed outer spine; b) the null-point-reconnected topology with the outer spine opened; c) formation of the flux rope and its related separatrices due to flare reconnection; d) merging of the separatrices and the formation of a transient complex topology; e) splitting of the merged separatrices to form two new open-spine null-point topologies; and f) transfer of magnetic shear from the closed flux rope to the open interplanetary field.
}
  \label{f-new topo}
\end{figure}

\noindent We evaluated the potential numerically using numerical fits to the elliptic integrals given by \citet{Milne-Thompson72}, and then derived the magnetic field from ${\bf B}({\bf r})={\bf \nabla} \times {\bf A}({\bf r})$. By superposing the solar background dipole field and the magnetic field created by the dipole ring (Eq.~\ref{eq-Aphi3}), for the parameter values specified in \S \ref{topology}, we obtained the 2.5D axisymmetric quadrupolar magnetic configuration shown in Figure \ref{f-topology}a. It is straightforward to show from the above expression that, for the special case $b = 0$, $a = R_{\odot} - d$, and $d \ll R_{\odot}$, $\vert {\bf B} \vert = B_0$ at the solar equator ($z = 0$, $\xi = 
R_{\odot}$), whether the dipole moment is tangent to the surface ($\mu_z = 1$) or normal to the surface ($\mu_\xi = 1$). This result motivated the particular form chosen for the 
normalization constant in Equation \ref{eq-Aphi1}.


\newpage

\newpage

\newpage

\newpage

\end{document}